\documentclass[useAMS,usenatbib,usegraphicx]{mn2e}
\usepackage{natbib}
\usepackage{amsmath}
\usepackage{amssymb}
\usepackage{mathtools}
\usepackage{graphicx}
\newcommand{\be}{\begin{equation}}
\newcommand{\ee}{\end{equation}}
\newcommand{\beq}{\begin{eqnarray}}
\newcommand{\eeq}{\end{eqnarray}}
\newcommand{\bear}{\begin{eqnarray}}
\newcommand{\eear}{\end{eqnarray}}
\newcommand{\ba}{\begin{array}}
\newcommand{\ea}{\end{array}}

\title{Detecting gravitational waves from mountains on neutron stars in the Advanced Detector Era}
\author[B.Haskell et al.] {B.~Haskell$^1$,  M.~Priymak$^1$, A.~Patruno$^{2,3}$, M.~Oppenoorth$^4$, A.~Melatos$^1$, P.~D.~Lasky$^{1,5}$\\
$1$ School of Physics, The University of Melbourne, Parkville VIC 3010, Australia\\
$2$ Leiden Observatory, Leiden University, PO Box 9513, 2300 RA Leiden, the Netherlands\\
$3$ ASTRON, the Netherlands Institute for Radio Astronomy, Postbus 2, 7990 AA Dwingeloo, the Netherlands\\
$4$ Copernicus Institute of Sustainable Development, University of Utrecht, P.O. Box 80.115, 3508 TA Utrecht, the Netherlands\\
$5$ Monash Centre for Astrophysics, School of Physics and Astronomy, Monash University, VIC 3800, Australia.}

\begin{document}
\maketitle
\begin{abstract} 

Rapidly rotating Neutron Stars (NSs) in Low Mass X-ray Binaries (LMXBs) are thought to be interesting sources of Gravitational Waves (GWs) for current and next generation ground based detectors, such as Advanced LIGO and the Einstein Telescope. The main reason is that many of the NS in these systems appear to be spinning well below their Keplerian breakup frequency, and it has been suggested that torques associated with GW emission may be setting the observed spin period. This assumption has been used extensively in the literature to assess the strength of the likely gravitational wave signal. There is now, however, a significant amount of theoretical and observation work that suggests that this may not be the case, and that GW emission is unlikely to be setting the spin equilibrium period in many systems. In this paper we take a different starting point and predict the GW signal strength for two physical mechanisms that are likely to be at work in LMXBs: crustal mountains due to thermal asymmetries and magnetically confined mountains. We find that thermal crustal mountains in transient LMXBs are unlikely to lead to detectable GW emission, while persistent systems are good candidates for detection by Advanced LIGO and by the Einstein Telescope. Detection prospects are pessimistic for the magnetic mountain case, unless the NS has a buried magnetic field of $B\approx 10^{12}$ G, well above the typically inferred exterior dipole fields of these objects. Nevertheless, if a system were to be detected by a GW observatory, cyclotron resonant scattering features in the X-ray emission could be used to distinguish between the two different scenarios.
\end{abstract}


\section{Introduction}

Rapidly rotating Neutron Stars (NSs) are considered an interesting source of Gravitational Waves (GWs) and are one of the main targets for current searches with ground based detectors, such as Virgo and LIGO \citep{riles13}. The characteristic amplitude of the GW signal scales with the square of the rotation frequency, thus making the more rapidly rotating NSs ideal candidates for detection.
In particular some of the most promising targets are likely to be accreting NSs in Low Mass X-ray Binaries (LMXBs). Not only are these NSs rotating with millisecond periods, but the process of accretion from the companion star can drive the growth of a quadrupolar deformation. Plausible mechanisms that may be at work are the creation of a ``mountain'' (i.e. any kind of non-axisymmetric deformation that gives rise to an $l=m=2$ mass quadrupole) supported by the elastic crust \citep{Bildsten98, UCB, mountain,BenMount} or by a solid core of exotic matter \citep{OwenQ, CFLm}, unstable modes of oscillation of the star  \citep{Nils98,Nils99} and magnetically supported mountains \citep{Cutler,HaskellM, Melatos, Max1,VIG09a}.

LMXBs were originally invoked as a source of GWs to solve an observational puzzle. In an LMXB the NS is spun up by matter accreted from the companion via a disc. This is, in fact, how old NSs are thought to be recycled to millisecond periods and eventually produce a millisecond radio pulsar after accretion stops \citep{AlpR,Rad1,swing}. One would therefore expect the NS to be spun up to its centrifugal break-up frequency, which is equation of state dependent, but generally well above 1 KHz \citep{Cook, Haensel}. This is not, however, what is observed. The distribution of spins in both LMXBs and millisecond radio pulsars appears to have a statistically significant cutoff at around 730 Hz  \citep{Chak, Patruno1}.

It is natural to ask what physical process removes angular momentum from the NS and prevents it from spinning up further. The first and most obvious candidate is the interaction between the stellar magnetic field and the accretion disc. This possibility was examined in detail by \citet{WZ} who found that, at least for the data available at the time, this scenario would involve an unexpected correlation between the accretion rate and magnetic field strength (which would also need to be higher than expected). This led to the alternative suggestion that GWs may be providing the torque needed to balance the accretion torques, and set the spin equilibrium period of these systems \citep{PP, Wagoner, Bildsten98}.

A corollary of GW torque-balance is that the brightest X-ray sources should also be the loudest GW emitters \citep{Bildsten98}.  This describes the nearby LMXB Scorpius X-1, which has been the subject of a number of LIGO and Virgo searches \citep{abbott07a, abbott07b, abadie11, aasi14} that have led to a 90$\%$ confidence upper limit for the gravitational wave strain of $h_{\rm rms}\approx10^{-25}$ around 150 Hz.  With advanced detectors now coming on line there is a strong case to develop directed data analysis algorithms \citep{sammut14} and all-sky pipelines that search for unknown binary systems \citep{goetz11}.

Although GW searches with initial LIGO are still not sensitive enough to probe the predictions of the GW torque balance scenario, the problem has been recently reassessed by several authors. 
\citet{ABCPHD} found that with current data the strong correlation between magnetic field and accretion rate found by \citet{WZ} is no longer needed and the measured spin period of most systems can be understood in terms of the disc/magnetosphere interaction \citep{A1}. Furthermore a detailed analysis of individual systems shows that many of them do, in fact, appear to be close to a propeller phase in which the spin-up torque is much weaker than in standard accretion models \citep{BrynAle,IGR}. Finally the measurements of spins and surface temperatures for most NSs in LMXBs are not consistent with theoretical predictions for GW emission due to an unstable r-mode (or at least not at a level that would allow for spin-equilibrium due to torque balance) \citep{windowW, HasDegHo, SiminR}

\begin{table*}
\begin{minipage}{150mm}
\begin{flushleft}
\begin{tabular}{l l l l l l l l l l l }
\hline
\hline
Source& &$\nu$ & &  $d$ & & $\langle \dot{M}\rangle$   & &  $\Delta t$& &  Ref.\\

      & & (Hz) & &  (kpc)   & & ($10^{-10}\rm\,M_\odot\,yr^{-1}$) &  &  (d)     & & \\
\hline
SAX J1808.4--3658 & &401& &  3.5 & &  4 & &  30 &  & \citet{t1}\\
XTE J1751--305 && 435 & &    7.5 &&     10 &&    10& & \citet{t2}\\
XTE J1814--338 && 314 &&    8   &&    2  && 60&& this work \\
IGR J00291+5934 && 599  &&  5     &&  6  && 14&& \citet{t3}\\
HETE J1900.1--2455 && 377   &&  5     &&  8  && 3000&&  \citet{t4}\\
Aql X-1 && 550    && 5      &&  10   && 30&&  \citet{t5}\\
Swift J1756.9--2508 & & 182.1  &&  8     &&  5  && 10&& \citet{t6}\\
NGC 6440 X-2 & & 204.8  &&  8.5    && 1  && 4&& this work \\
IGR J17511--3057 && 244.9   && 6.9    && 6   && 24&&\citet{t7} \\
IGR J17498--2921 & & 400.9   && 7.6     &&  6  && 40& & \citet{t8}\\
Swift J1749.4-2807 && 518     && 6.7     && 2  && 20& & \citet{t9}\\
\hline
EXO 0748--676 & & 552     && 5.9     && 3  && 8760& & \citet{t10} \\
4U 1608--52 && 620     && 3.6     && 20  && 700&& \citet{t11}\\
KS 1731--260 && 526     && 7       && 11    && 4563& & \citet{t12}\\
SAX J1750.8--2900 & & 601     && 6.8     && 4   && 100& & this work\\
4U 1636--536 && 581   &&  5     &&  30 && pers. & & this work\\
4U 1728--34 && 363    && 5      && 5 &&	pers. & & \citet{t13}\\
4U 1702--429 && 329    && 5.5    && 23 && pers. & & this work \\
4U 0614+091 && 415    && 3.2    && 6 && pers. && \citet{t14}\\
\hline
\end{tabular}
\end{flushleft}
\caption{LMXBs for which we have obtained an estimate of the outburst duration $\Delta t$ and average accretion rate $\langle \dot{M}\rangle$. Where the reference column indicates `this work', we have used a fiducial power law index of $\Gamma=2$ and the galactic absorption column
from \citet{Kalberla05}. We also list the distance $d$ of the system and the spin frequency $\nu$. Sources in the top half of the table are AMXPs, while those in the bottom half are nuclear powered pulsars and their frequency is inferred from the frequency of burst oscillations, as explained in the text. We do not attempt to explicitly estimate the errors associated with these measurements. The most uncertain quantity is the distance, but our main conclusions on the detectability of the GW signals are unlikely to change unless there is a substantial error in the values above.}
\end{minipage}
\label{aletable}
\end{table*}

GW torque balance supplies a useful upper limit to calibrate searches. However if it is not the driving force behind pulsar spin evolution, it is natural to ask at what level the physical mechanisms mentioned above will give rise to GW emission, and whether it is likely to be detected. This question is crucial, given that \citet{Watts08} showed that even at the torque balance level these systems would be challenging to detect. In this paper we explore the non-torque-balance scenario in more detail. We focus on ``mountains'',  supported either by elasticity or magnetic stresses, and discuss the level at which GW emission may be expected. We also take the discussion one step further and ask, given a GW detection, what constraints can be set on NS interior physics and how one could distinguish between the different mechanisms giving rise to the mountain using electromagnetic (e.g. X-ray) observations.

\section{Thermal mountains}

\subsection{Crustal heating}

The outer, low density layers of a NS are thought to form a crystalline crust of ions arranged in a Body Centred Cubic (BCC) lattice [although recent work by \citet{Peth} suggests that much more inhomogeneous configurations may be possible]. Above densities of $\approx 10^{11}$ g/cm$^3$, neutrons drip out of nuclei and form a superfluid in mature NSs with internal temperatures $T\lesssim 10^9$ K. At higher densities several phase transitions may occur, with nuclei no longer being spherical but forming rods and plates, the so called ``pasta" phases \citep{Ravenhall}, until at $\approx 2 \times 10^{14}$ g/cm$^3$ there is a transition to a fluid of neutrons, protons and electrons which forms the core of the NS.

In LMXBs accreted matter, composed of light elements, is buried by accretion and compressed to higher densities, where it undergoes a series of nuclear reactions such as electron captures, neutron emission and pycnonuclear reactions \citep{haenselR}. The observed cooling of transient LMXBs, as they enter quiescence, is consistent with a crust that has previously been heated by such reactions [see e.g. \citet{trans} and references therein, although not all details of the cooling processes are fully understood \citep{NatN, SNat}].

Accretion asymmetries can produce asymmetries in composition and in heating, which in turn deform the star and lead to a quadrupole \citep{UCB}. 
Once the quadruple $Q_{22}$ is known the GW amplitude can be calculated as:
\be
h_0=\frac{16}{5} \left(\frac{\pi}{3}\right)^{1/2}\frac{G Q_{22}\Omega^2}{d c^4},\label{h}
\ee
where $G$ and $c$ are the gravitational constant and the speed of light respectively, $d$ is the distance to the source and $\Omega$ is the angular frequency of the star. Note that we are considering a quadrupolar $Y_{22}$ deformation, as this harmonic dominates GW emission. In this case GWs are emitted at twice the rotation frequency of the star.
An approximate expression for the quadrupole due to asymmetric crustal heating from nuclear reactions in the crust is given by \citep{UCB}:
\be
Q_{22}\approx 1.3 \times 10^{35} R_6^4 \left(\frac{\delta T_q}{10^5 K}\right) \left(\frac{Q}{30\mbox{MeV}}\right)^3 \mbox{g cm$^2$},\label{quad1}
\ee
where $R_6$ is the stellar radius in units of $10^6$ cm, $\delta T_q$ is the {\it quadrupolar} component of the temperature variation due to nuclear reactions and $Q$ is the reaction threshold energy. Higher threshold energies correspond to higher densities.
In general the reactions will heat the region by an amount \citep{Rutledge}:
\be
\frac{\delta T}{(10^6 K)}\approx C_k^{-1} p_d^{-1}Q_n\Delta M_{22},\label{heating}
\ee
where $C_k$ is the heat capacity per baryon in units of the Boltzman constant $k_B$, $p_d$ is the pressure, in units of $10^{30}$ erg cm$^{-3}$, at which the reaction occurs, $Q_n$ is the heat per unit baryon (in MeV) deposited by the reactions and $\Delta M_{22}$ is the deposited mass in units of $10^{22}$ g.
Note that $\delta T$ in (\ref{heating}) is the total increase in temperature; only a fraction $\delta T_q/\delta T\ll 1$ is likely to be asymmetric in general and specifically quadrupolar. \citet{UCB} estimate that $\delta T_q/\delta T\leq 0.1$, but in reality the ratio is poorly known.

After an accretion outburst, as the system returns to quiescence, the deformations are erased on the crust's thermal timescale \citep{B98}:
\be
\tau_{th}\approx 0.2 \;p_d^{3/4}\; \mbox{yr}
\label{tscale}
\ee
If the system is in quiescence for longer than the thermal timescale in (\ref{tscale}), $Q_{22}$ is likely to be washed out and a new mountain is rebuilt during the next outburst. A shorter recurrence time, on the other hand, could lead to an incremental accumulation of material.
However, compositional asymmetries may be frozen into the crust, and not be erased on a thermal timescale, allowing for the mountain to be built incrementally \citep{UCB}. This scenario predicts the formation of large quadrupoles in all transient systems ($10^{38}\lesssim Q\lesssim10^{40}$ g cm$^2$), as we discuss in section \ref{trans}. The implied spin-down rate, in the case of four transient systems (SAX J1808.4--3658, XTE J1751--305, IGR J00291+5934 and SWIFT J1756.9--2508) is already excluded by measurements of the spin-down rate between outbursts \citep{review}. We do not consider this scenario further, but note that if it were to occur in any transient system, the GW strain would be comparable to that of a persistent system.

\subsection{Maximum quadrupole}

Large stresses can break the crust, so one should also ask how large a mountain the star can sustain. This problem has been studied by different authors in Newtonian physics \citep{UCB,mountain} and, more recently, in general relativity \citep{BenMount}. The results depend critically on the breaking strain $\bar{\sigma}_c$ of the crust, i.e. the average strain $\bar{\sigma}=\bar{T}/\mu$ that can be built up before the crust cracks, where $\mu$ is the shear modulus of the crust and $\bar{T}$ the average stress. The breaking strain of a NS crust is not well known, but is known to be $\bar{\sigma}_c\approx 10^{-2}$ for perfect crystals in a laboratory setting, and recent molecular dynamics simulations have shown that it may reach  $\bar{\sigma}_c\approx 10^{-1}$ in NS crusts \citep{chuck}.
The maximum quadrupoles that can be sustained are thus of the the order of $Q_{22}\approx 10^{38}-10^{39}$ g cm$^2$ for more massive stars ($M\approx 2 M_\odot$) and  $Q_{22}\approx 10^{39}-10^{40}$ g cm$^2$ for less massive stars ($M\approx 1.2 M_\odot$), depending on the exact equation of state.

\subsection{Gravitational radiation}

It is natural to ask, for the currently known LMXBs, how large a thermal mountain can grow and if it is detectable by current and next generation interferometers, such as Advanced LIGO or the Einstein Telescope (ET).
To answer these questions let us examine the LMXBs whose spins are known. They can be divided in two classes: the Accreting Millisecond X-ray Pulsars (AMXPs), which are detected as pulsars and can thus be timed, and the Nuclear Powered (NP) pulsars which do not pulsate, but exhibit quasi periodic oscillations in the tails of  type II nuclear bursts. The frequency of these oscillations is a measure of the spin period, as confirmed by observations of burst oscillations in sources that are also detected as X-ray pulsars \citep{review}. The details of the LMXBs we use are presented in table \ref{aletable}. The other quantities listed in table \ref{aletable} are the distance to the source and additionally the average duration $\Delta t$ and average mass accretion rate $<\dot{M}>$ during outbursts. The amount of mass that is accreted during an outburst can then be obtained as $\Delta M=<\dot{M}>\Delta t$, and inserted into equations (\ref{heating}) to calculate the temperature increase due to nuclear reactions in the crust.

To calculate the average mass accretion rate for all sources we followed
two different approaches. For AMXPs we used the data collected by the
\textit{Rossi X-ray Timing Explorer} (\textit{RXTE}) and recorded with
the Proportional Counter Array (PCA; see \citealt{jahoda2006}).  We
used the Standard-2 data mode and extracted the 2--16 keV X-ray fluxes
for all outbursts, following the procedure by van Straaten et al. (2003).
The fluxes for each outburst were averaged
assuming a fiducial spectral index (i.e. assuming that the spectral index remains constant during the course of the
outburst and between different outbursts) taken from the literature.
We then extrapolated to the 0.1-100 keV (bolometric) flux.  We used the
unabsorbed luminosity (where we take the hydrogen absorption column
$N_{H}$ reported in the literature, see table \ref{aletable}). The bolometric
luminosity was then calculated from the distance in table
\ref{aletable} and the mass accretion rate as given by
$L_{acc} = G\,M\,\dot{M}\,R = \dot{M}\eta\,c^{2}$.  Here, $L_{acc}$ is
the bolometric accretion luminosity, and we assumed a mass of $M=1.4\,M_{\odot}$, and a radius $R=10$ km and $\eta$ is the conversion efficiency for
the rest-mass into energy.  After calculating the average mass
accretion rate for each outburst we selected (and reported in table \ref{aletable})
the highest value obtained (i.e., we consider the biggest
possible mountain).

For the nuclear-powered accreting pulsars we used instead data from the All-Sky
Monitor (ASM) onboard \textit{RXTE}, which operated in the 1.3--12.1
keV band. We used the ASM rather than the PCA because all the 8 sources
selected are either persistent sources or have long outbursts. The
ASM, being a monitoring instrument, has a much better data coverage
(although with lower sensitivity and a narrower energy band).  In
this case we selected the absorption column $N_{H}$ and spectral index
$\Gamma$ from the literature (whenever available) or, when no spectral analysis was available, used the
galactic $N_{H}$ and a simple power law model with spectral index $\Gamma=2$. 

We caution that the results may suffer from systematic errors in both distance $d$ and spectral index $\Gamma$. However the estimates are likely to be sufficiently accurate
for our purposes. The main conclusions of this paper will not
change unless there is a substantial error in our assumptions that can
change the mass accretion rate by orders of magnitude (e.g., a substantial
error in the distance). 

\begin{figure*}
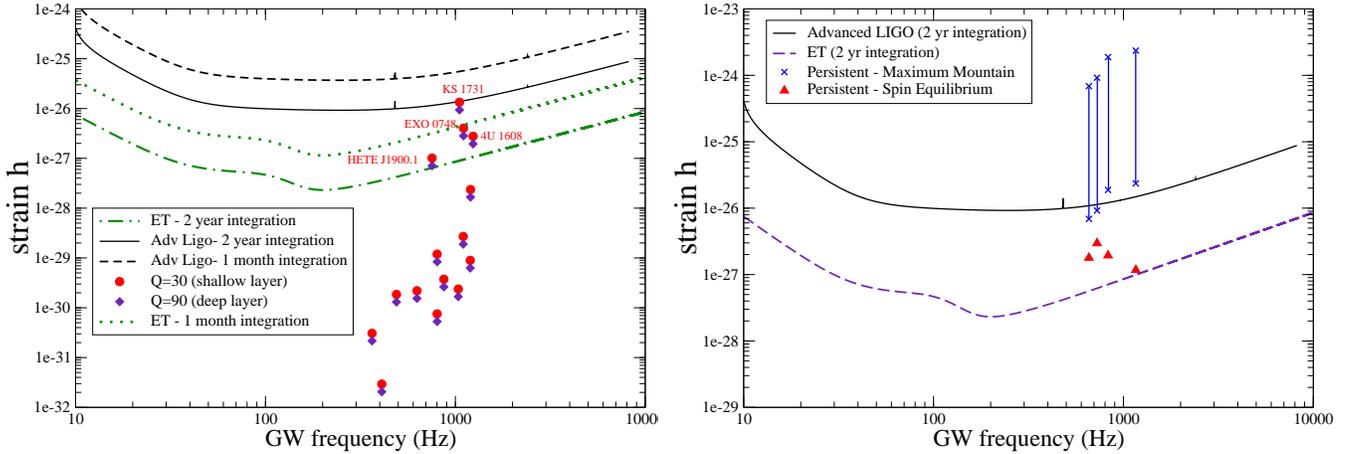

\centerline{\includegraphics[width=8.8cm]{figures/nonpersistET.eps}\;\includegraphics[width=8.8cm]{figures/persistET.eps}}
\caption{Gravitational wave strain vs frequency for mountains in AMXPs and NP pulsars. In the left panel we show transient sources, for which the mountain is the largest that can be created during an outburst, both in the case of a shallow ($Q=30$ MeV) and of a deep ($Q=90$ MeV) capture layer.  In the right panel we show the persistent sources, for which we assume both the maximum mountain the crust can sustain (crosses), and a mountain that would give spin equilibrium from torque balance (solid triangles). The bars indicate the range given by uncertainties on the breaking strain, as described in the text.}\label{ligofigs}
\end{figure*}

\subsection{Transient sources}
\label{trans}
In the left panel of figure \ref{ligofigs} we show the GW strain corresponding to the maximum
mountain that could be created during an outburst [equations (\ref{quad1}) and (\ref{h})], assuming that
$\delta T_q/\delta T=0.1$. We consider this to be a reasonable upper limit, as a
significantly larger fraction $\delta T_q/\delta T$ would lead to detectable pulsations in
quiescence for some of the sources, as we shall see in section \ref{therm}. Note, however, that there is currently no physically motivated estimate for 
$\delta T_q/\delta T$, and the true value may be much smaller. We consider two capture layers, a shallow one close to
neutron drip (where most of the heat is predicted to be deposited
\citep{haenselR}) with a threshold of $Q=30$ MeV, and a deeper layer at a pressure of $p=10^{32}$ dyne/cm$^2$, with a threshold
energy of $Q=90$ MeV. All values lie below the maximum quadrupole
that the crust can sustain. Note also that the increase in the quadrupole
$Q_{22}$ due to the higher threshold energy $Q$ is more
than offset by the decrease in heating at higher pressures, as
obtained from eq. (\ref{heating}). The results for the deep and shallow capture layers are thus very similar.
The thermal timescale for the deeper capture layers is, however,  $\tau_{th}\approx 6$ years. Hence a `deep' mountain may never relax entirely in systems such as Aql X-1 that have frequent outbursts,  with recurrence times shorter than $\tau_{th}$. These systems may effectively behave as persistent sources for our purposes, and harbour larger mountains.

In the left panel of figure (\ref{ligofigs}) we compare our results to the sensitivity achieved by Advanced LIGO (assuming both detectors have the same sensitivity) and ET, first by assuming an integration time of 1 month (an average duration for an outburst) and then of 2 years. It is quite clear from the figure that, even for a 2 year integration, most systems fall well below the sensitivity curve. 
Strain sensitivity curves for Advanced LIGO and ET are respectively taken from the public LIGO document\footnote{https://dcc.ligo.org/LIGO-T0900288/public} and \citet{hild08}.  A fully coherent search over time, $T_{\rm obs}$, is sensitive to a strain of 
\be
h\approx11.4\sqrt{S_n(\nu)/T_{\rm obs}},\label{11}
\ee
 where $S_n(\nu)$ is the detector noise power spectral density, and the factor 11.4 accounts for a single trial false alarm rate of 1\% and a false dismissal rate of 10\% \citep{abbott07a, Watts08}.

It is unlikely that transient systems will be strong enough sources for Advanced LIGO, but they are promising sources for ET. This is essentially the same conclusion of \citet{Watts08}, who considered emission at the torque balance level, which is higher than the strain we calculate [note that both our estimates and those of \citet{Watts08} assume mountains that are smaller than the maximum that the crust can sustain before breaking].
A few systems appear to be close to the threshold for detection. However these systems are unlikely to be good targets for upcoming GW searches, as they have all just entered quiescence after long outbursts, during which large amounts of mass were accreted and the crust was heated considerably. The mountain is currently relaxing on a timescale $\tau_{th}$ and the recurrence time between accretion outbursts is likely to be long. It is thus probable that they will not be `on' as continuous GW sources during Advanced LIGO observations.

\subsection{Persistent sources}

For the persistently accreting sources the situation is different. We assume that
ongoing accretion builds the largest mountain that can be
sustained. We take the quadrupole to be in the range $10^{38} \mbox{g cm$^2$}\lesssim Q_{22}\lesssim 10^{40} \mbox{g cm$^2$}$,
to account for the uncertainty in mass and equation of state, as
estimated by \citet{BenMount}.  The results are shown in the right hand panel of figure
\ref{ligofigs}. The error bars account for the range discussed
above. We also present the torque balance upper limits on
$Q_{22}$, as in \citet{Watts08}. The results for the maximum mountain
comfortably exceed the torque
balance limits. If accretion is ongoing, the quadrupole can thus become
larger than the value needed for torque balance. In this
scenario there is thus a net spin-down torque due to GW emission, and the prediction for the spin-down rate is:
 \be
\dot{\nu}\approx 6\times 10^{-13}
\left(\frac{\nu}{500\mbox{Hz}}\right)^5\left(\frac{Q_{22}}{10^{38}\mbox{g
    cm$^2$}}\right)\; \mbox{Hz/s},\label{spindown} \ee (where we have
assumed a moment of inertia $I=10^{45}$ g cm$^2$ for the star). 
Such spin-down is sufficiently strong 
to be detectable with current instrumentation. However, none of the
 persistent sources considered here have ever shown accretion powered
pulsations that would allow us to test this prediction. Continued deep searches for pulsations from these objects is thus of significant importance for GW science.

Another issue to consider is the amount of internal heating required to sustain a large quadrupole. Rearranging (\ref{heating}) we
see that, for a fiducial star of radius $R=12$ km, one has:
 \be \delta
T_{q}\approx 3\times 10^7 \left(\frac{Q_{22}}{10^{38}\mbox{g
    cm$^2$}}\right)\left(\frac{30\mbox{MeV}}{Q}\right)^3\; \mbox{K},
     \ee
For hot sources with internal
temperatures $T=10^8$ K, high values of the quadrupole (around
$Q_{22}\approx 10^{39}$ g cm$^2$) would require $\delta T_{q}/T>1$, even for
deeper capture layers. Such high values of $\delta T_{q}/T>1$ would lead to pulsations in quiescence at a level that is not observed. However for lower values of $Q_{22}$, deeper
capture layers and higher activation energy Q, the temperature
perturbation is $\delta T_{q}/T\leq 0.1$. During accretion outbursts
the resulting perturbations to the luminosity are $\delta L_{bol}\lesssim 10^{32}\rm\,erg\,s^{-1}$, and are not
visible \citep{UCB}, as the emission is at much higher levels
  ($L_{bol}=L_{acc}\simeq 10^{35}-10^{37}\rm\,erg\,s^{-1}$). However such levels
of heating can make the quiescent flux vary, as we argue below.

\begin{figure}
\centerline{\includegraphics[width=9cm]{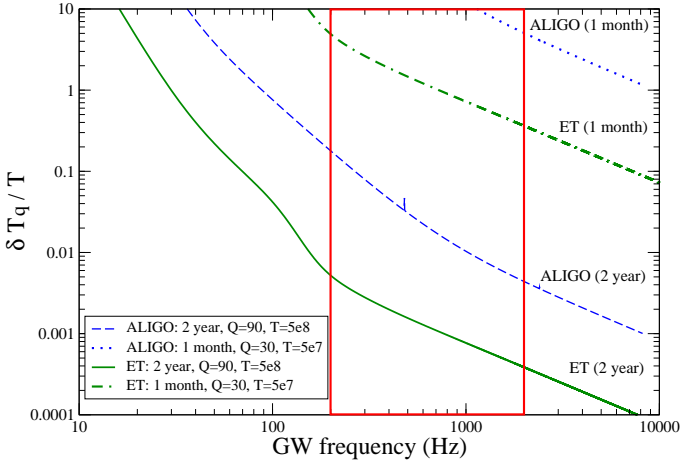}}
\caption{Sensitivity of current and next generation GW detectors to gravitational waves sourced by quadrupolar temperature deformations $\delta T_q$ in deep ($Q=90$ MeV) and shallow ($Q=30$ MeV) layers of the NS crust. The gravitational wave strain is expressed in terms of the temperature perturbations $\delta T_q/T$ that give rise to the mountain, as described in the text. Both 2nd and 3rd generation detectors will probe regimes of physical interest, with Advanced LIGO probing the $\delta T_q/T\approx 0.1$ regime, and ET the $\delta T_q/T\approx 0.01$ regime. We show the sensitivity of Advanced LIGO both for a 1 month integration, $Q=30$ MeV, and background temperatures $T=5\times 10^7$ K (dotted curve), corresponding to the case of a short outburst, and for a 2 year integration, $Q=90$ MeV and a background temperature of $T=5\times 10^8$ K (dashed curve), more appropriate for a persistent system. Similarly we show the sensitivity of ET for $T=5\times 10^7$ K, $Q=30$ MeV, and a 1 month integration (dot-dashed curve) and $T=5\times 10^8$ K, $Q=90$ MeV, and a 2 year integration (solid line curve). The region enclosed by the red box  is that most relevant for LMXBs.}\label{heatfig}
\end{figure}

We can summarise the discussion above by asking what a GW detection implies for deep
crustal heating. We use equations (\ref{h}) and
(\ref{quad1}) to represent the sensitivity curve of Advanced Ligo and
ET in terms of an equivalent quadrupolar temperature deformation $\delta T_q$, as shown in
figure \ref{heatfig}. We consider two fiducial systems
at a distance $d=5$ kpc: a system that undergoes shorter outbursts and
is colder ($T=5\times 10^7$ K), for which we integrate the GW signal
over the fiducial duration of the outburst (1 month); and a hotter
($T=5\times 10^8$ K), persistent system for which we integrate the GW
signal over a 2 year period. Figure \ref{heatfig} shows that Advanced LIGO and ET will
probe the $\delta T_q/T\approx 0.1$ regime, with ET probing the
possibly more realistic $\delta T_q/T\leq 0.01$ regime. This is also the
order of magnitude of the perturbations expected in the quiescent flux
(see section \ref{therm}), which may be detectable with future X-ray
satellites such as the Large Observatory for X-ray Timing (LOFT) or the Neutron star Interior Composition Explorer (NICER).

In the analysis above we make many approximations. First and foremost we only consider two capture layers in the stars. In reality all layers contribute to $Q_{22}$, leading to larger quadrupoles than those discussed above \citep{UCB}. For shorter outbursts the reduced heating at higher densities offsets the higher quadrupole $Q_{22}$ in those regions. The reactions that deposit the most heat thus dominate, independently of density. We are accounting for what is considered to be the most important layer at neutron drip \citep{haenselR}, so unless there is significantly more heating in deeper layers that previous calculations have not accounted for, it is unlikely that our results severely underestimate $Q_{22}$.
In general the result of our analysis is that thermal mountains on NSs in transient LMXBs are likely to be very challenging to detect, even with third generation detectors. Persistent systems, however, offer a promising target and electromagnetic observations may allow further constraints on the physics of the system, as we shall see in section \ref{dist}.

\subsection{X-ray flux variations}

\label{therm}

We now focus on the observable X-ray flux variations induced by a thermal perturbation due to a mountain in the
crust. As already discussed we restrict our attention to perturbations of the thermal quiescent emission. We thus assume that a mountain has been created during an accretion outburst, and study how the associated thermal perturbations evolve as the system returns to quiescence.  To understand how the surface flux is
affected we consider the quadrupolar flux variations as linear
perturbations on a spherically symmetric background. We start by obtaining the spherical background model for the temperature profile from the Newtonian heat transport equations
in the crust: \be C_v\frac{\partial T}{\partial
  t}=\nabla({K}\nabla T)-\rho\epsilon\label{bk}, \ee where $T$
is the temperature, $C_v$ the heat capacity, ${K}$ the conductivity,
$\rho$ the density and $\epsilon=\epsilon_\nu-\epsilon_h$, with
$\epsilon_\nu$ the neutrino emissivity and $\epsilon_h$ the energy
deposition rate.  To simplify our treatment and make a first
assessment of detectability, we will use an $n=1$ polytrope for the
equation of state, and analytic expressions for the
contribution of electrons in the crust to the conductivity
\citep{FI81} and specific heat \citep{Max79}, from which one obtains: \beq
{K}&=&10^{16}\left(\frac{\rho}{10^6\mbox{g/cm$^3$}}\right)^{1/3}\left(\frac{T}{10^8
  \mbox{K}}\right)\;\mbox{erg/(cm s K)},\label{kk}\\
   C_v&=&3.72\times
10^{17}\left(\frac{\rho}{\rho_0}\right)^{4/3}\left(\frac{T}{10^8
  \mbox{K}}\right)\;\mbox{erg/(cm$^3$ K)},
   \eeq where $\rho_0$ is the nuclear saturation
density. We set $\epsilon_h=0$ (note this is true for the background, but for the perturbations we will have $\delta\epsilon_h\neq 0$), and for the neutrino
emissivity we approximate the results of \citet{HKY96} for $\nu\bar{\nu}$ electron Brehmstrahlung as: \be
\epsilon_\nu=6.46\times
10^{18}\left(\frac{\rho}{10^{12}\mbox{g/cm$^3$}}\right)\left(\frac{T}{10^9
  \mbox{K}}\right)^6\;\mbox{cm$^2$/s$^3$} \ee
   At the boundary with the core we assume a constant temperature and for the outer boundary we assume that the emission from the surface is
thermal, i.e. $-K\nabla T=(R^2/R_*^2) \sigma T_s^4$, with $\sigma$ the
Stefan-Boltzman constant. The stellar radius is $R$, and $R_*$ is the
radius at which we fix the outer boundary of our numerical grid. The
surface temperature $T_s$ at $R$ is then obtained
from the temperature $T$ at $R_*$ using the prescription of \citet{Gudmundsson}: \be
\left(\frac{T_s}{10^6\mbox{K}}\right)=g_{14}\left(18.1\frac{T}{10^9\mbox{K}}\right)^{2.42},\label{gud}
\ee with $g_{14}$ the gravitational acceleration in units of $10^{14}$
cm s$^{-2}$. Note that one can model the composition of the outer
layers in more detail (see e.g. \citealt{HasDegHo} and
\citealt{SiminR}). However, given the many simplifying assumptions and the uncertainties associated with the measurements in table \ref{aletable}, we use the expression in (\ref{gud}), as it is unlikely to be the main source of error in our analysis.

We obtain our background model by specifying a core temperature at the inner boundary (the crust/core interface) and evolving equation (\ref{bk}) until we obtain an equilibrium. We are now ready to evolve the quadrupolar temperature perturbations due to the mountain on this background.
The evolution equation for an $l=m=2$ perturbation takes
the form \citep{UCB}: \beq C_v\frac{\partial\delta T_q}{\partial
  t}&=&-\frac{1}{r^2}\frac{\partial}{\partial r}\left(r^2 K
\frac{\partial\delta T_q}{\partial r}\right) -l(l+1) \frac{ K \delta
  T_q}{r^2}+\nonumber\\ &&\rho\epsilon\left(\frac{\delta
  K}{K}-\frac{\delta\epsilon}{\epsilon}\right)+F_Q\frac{\partial}{\partial
  r}\left(\frac{\delta K}{K}\right),\label{perturbo} \eeq with $F_Q=-K\nabla T$ the
background flux obtained from the equilibrium solution of
(\ref{bk}). We obtain $\delta K$ from equation (\ref{kk}) with the condition $\delta\rho=0$, as in \citep{UCB}. At the boundary with the core we assume that $\delta
T_q=0$, while at the outer boundary we perturb the thermal flux
condition, so that one has $\delta F=4 (R^2/r^2) \sigma T_s^3 \delta
T^s_q$, with \be \left(\frac{\delta T^s_q}{10^6\mbox{K}}\right)=2.42\;
g_{14}\left(18.1\frac{T}{10^9\mbox{K}}\right)^{2.42}\frac{\delta
  T_q}{T}. \ee 
We take $\delta\epsilon=\delta\epsilon_h$ and assume $\delta\epsilon_h$ to be due to quadrupolar energy deposition in the capture layers. For the deep capture layer, we specify an energy deposition term 
$\delta\epsilon_h$ with a Gaussian radial profile, located at a pressure of $P=10^{32}$ dyne/cm$^2$, and
with a half width of 5 m for the deep capture layer, and at
$P=10^{30}$ dyne/cm$^2$ and with a half width of 1 m for the
shallow layer. Evolving equation (\ref{perturbo}) we find that, as the problem is linear in the perturbations, to a very good
approximation the following relations hold: \beq &&\frac{\delta F}{F_Q}\approx 1.29\; \frac{\delta
  T_{q}}{T},\;\;\;\;\mbox{(deep layer)}\label{fraction1}\\ &&\frac{\delta
  F}{F_Q}\approx 1.48 \;\frac{\delta T_{q}}{T},\;\;\;\;\mbox{(shallow
  layer)}\label{fraction2} \eeq with very little dependence on the
chosen background temperature $T$. We remind the reader that we are normalising to the
quiescent (thermal) flux $F_Q$ obtained from the equilibrium solution of
(\ref{bk}).

\begin{figure}
\vspace{6mm}
\centerline{\includegraphics[width=9cm]{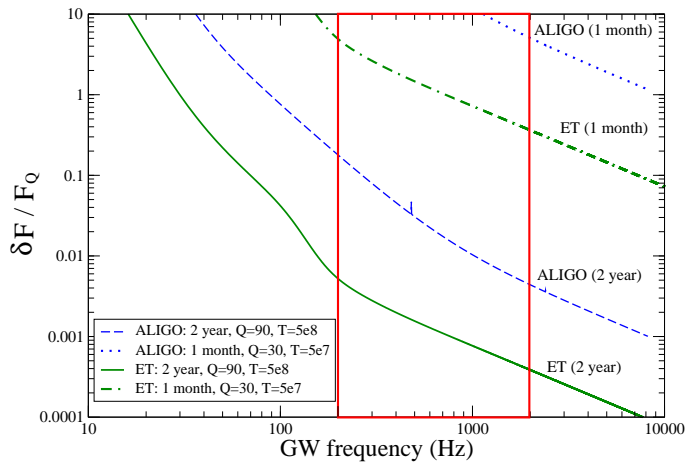}}
\caption{The pulsed fraction $\delta F/F_Q$ corresponding to a GW detection at threshold [as obtained from equation (\ref{11})] for Advanced LIGO and ET. For deep capture layers ($Q=90$ MeV) and a background temperature of $T=5\times 10^8$ K, we show the results for an integration time of 2 years (dashed curve for Advanced LIGO, solid curve for ET). Physically this is due to the fact that in the deep crust $\tau_{th}>2$ years, and the mountain will thus not relax significantly during the observation. For shallow capture layers ($Q=30$ MeV) and $T=5\times 10^7$ K, we show the results for an integration time of 1 months (dotted curve for Advanced LIGO, dot-dashed curve for ET). The region enclosed by the red box is the region of interest for LMXBs.}\label{heat2}
\end{figure}

In quiescence the quadrupolar temperature perturbations associated with a mountain and gravitational wave emission [equations (\ref{quad1}) and (\ref{heating})] thus perturb the X-ray flux from the surface [equations (\ref{fraction1}) and (\ref{fraction2})], and as the star rotates this leads to pulsations at twice the rotation frequency (i.e. the same frequency as the gravitational waves).
In figure \ref{heat2} we show the sensitivity curve for Advanced LIGO and ET in terms of an equivalent pulsed fraction of the X-ray flux. We can see that if it is possible to integrate the signal for 2 years (physically this corresponds to a capture layer deep enough that $\tau_{th}\gg 2$ years, and the mountain is not dissipated significantly during the observation), both Advanced LIGO and ET can probe an interesting region of parameter space, with $\delta F/F_Q\lesssim 0.01$.


\section{Magnetic mountains}
\label{magsec}
\subsection{Hydromagnetic evolution}
\begin{figure*}
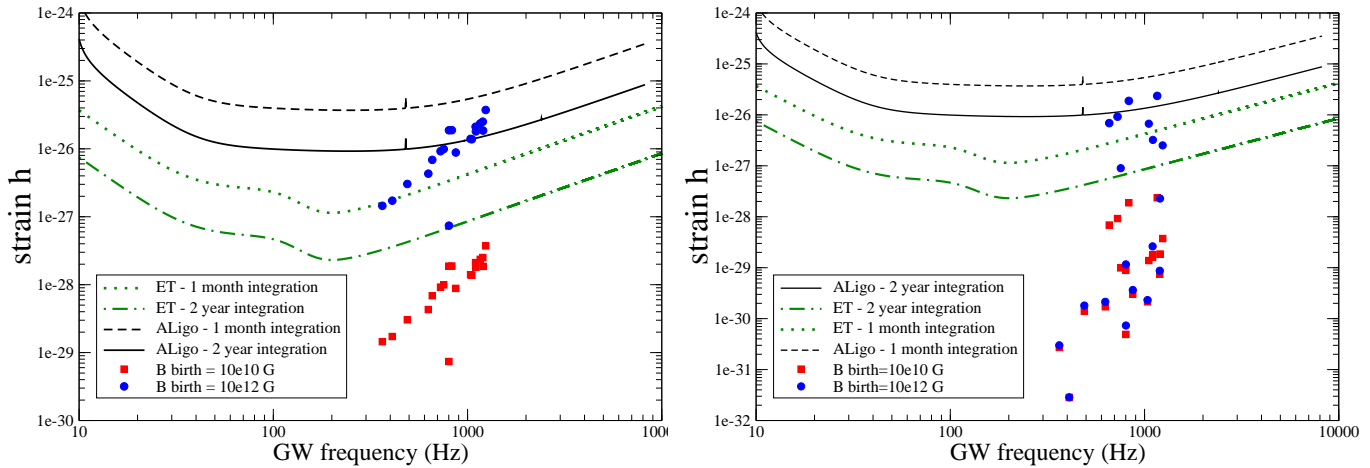

\centerline{\includegraphics[width=9cm]{figures/maximumET.eps}\includegraphics[width=9cm]{figures/transientmagET.eps}}
\caption{GW strain vs frequency for the systems in table \ref{aletable}, for two magnetic mountain scenarios. In the left panel we show the strain that can be achieved assuming that the magnetic mountain does not decay between outbursts, for two values of the magnetic field, $B_*=10^{10}$ G and $B_*=10^{12}$ G. In the right panel we consider the scenario in which the mountain decays between outbursts. Detection will be very challenging for both Advanced LIGO and ET, unless $B_*\approx10^{12}$ G before the onset of accretion.}\label{magfigs1}
\end{figure*}

Accretion not only perturbs the structure of the star by affecting nuclear reactions in the crust, but it also deforms the stellar magnetic field. As matter is accreted and spreads towards the equator it drags the field with it, and compresses it. This leads to a {\it locally} strong field that can sustain a ``magnetic'' mountain \citep{Andrew2,Melatos, VIG09}. This can lead to much larger deformations than those due to the overall background magnetic field, even if the internal toroidal component of the field is much stronger than the inferred external magnetic dipole \citep{RicNew}. Recent calculations have shown that  for realistic equations of state the mountain could lead to a detectable GW signal \citep{Max1}. Note also that magnetic mountains are not sustained by crustal rigidity and the resulting quadrupole can thus be larger than the value required to crack the crust.

One of the main differences with respect to thermal mountains is that the time-scale on which a magnetic deformation relaxes, after an outburst, is not the thermal timescale $\tau_{th}$, but the slower Ohmic dissipation timescale $\tau_o\geq 10^8$ years \citep{VIG09}. Hence a mountain forms gradually over several outbursts. Grad-Shafranov calculations indicate that the hydromagnetic structure of a mountain conforms to a single-parameter family of solutions which, once the size of the accreting polar cap is fixed, are function only of the mass accreted over the systems lifetime, $M_a$.
This suggests that magnetic mountains can be treated as the persistent sources of the previous section. The main difference is that the quadrupole does not depend on crustal rigidity, but on the magnetic field strength when accretion begins, which we denote $B_*$ (note that this is different from, and generally lower than, the expected NS magnetic field at birth, as obtained form population synthesis models \citep{Kaspi06}), the initial field structure, and $M_a$ \citep{Melatos, Andrew2, Max1}.

As more mass is accreted the external dipolar component of the field, $B_{ext}$, is quenched according to \citet{Shib}:
\be
B_{ext}=B_*\left(1+\frac{M_a}{M_c}\right)^{-1}\label{quench}
\ee
and the mass quadrupole is given by
\be
Q_{22}\approx 10^{45}\; A \left(\frac{M_a}{M_\odot}\right)\left(1+\frac{M_a}{M_c}\right)^{-1} \mbox{g cm$^2$} \label{quad}
\ee
where $A\approx 1$ is a geometric factor that depends on the equation of state and accretion geometry \citep{Melatos}, while $M_c$ is the critical amount of accreted matter at which the mechanism saturates, which also depends on the equation of state \citep{Max1}. The estimates above are valid to leading order in $M_a/M_c$; for $M_a\approx M_c$ they are no longer accurate and numerical solutions are necessary. General relations for the critical mass were derived by \citet{Melatos} and \citet{Andrew2} for isothermal mountains, while for more realistic equations of state [models C and E of \citet{Max1}], one has $M_c\approx 10^{-7} (B_*/10^{12} \mbox{G})^{4/3} M_\odot$. In the regime $M_a\gg M_c$, both relations are expected to deviate significantly from the simple estimates above in (\ref{quench}) and (\ref{quad}). Numerical simulations cannot probe this regime; instead one finds that, for $M_a\gtrsim M_c$ one has $0.01\leq B_{ext}/B_*\leq 0.1$ and quadrupoles in the range $10^{37}\lesssim Q_{22}\lesssim10^{38}$ g cm$^2$, for an initial field of $B_*=10^{12.5}$ G. Note, however, that the main difficulty in pushing the simulations to $M_a>10 M_c$ is numerical. The only firm upper limit on the suppression of the external dipole field come from Ohmic diffusion, which limits the burial of the field at a level of $B_{ext}/B_*\approx 10^{-4}$ \citep{VIG09}. 
\subsection{Pre-accretion magnetic field}

What limits can we set on $B_*$, the strength of the magnetic field at the onset of accretion?
Observational constraints can be obtained from measurements of the spin down between outbursts for 4 systems [see \cite{review} and references therein], which are consistent with $B_{ext}\approx 10^8$ G. The magnetic fields inferred for millisecond radio pulsars are also in the range $B_{ext}\approx 10^{8}$ G. Furthermore observations of a slow (11 Hz) pulsar in Terzan 5, IGR J17480-2446, indicate that this system, which is thought to have been accreting for a shorter period of time than most of the LMXB population, may have a stronger field $10^{9} \mbox{G}\lesssim B_{ext} \lesssim10^{10}$ G \citep{Cavecchi}. 
It is thus plausible that one starts with $B_*\gtrsim 10^{9}$ G, and that the external field is reduced to $B_{ext}\approx 10^8$ by accretion.  

If $B_*\gtrsim 10^{11}$ G,
and polar magnetic burial is very short lived, we would expect $B_{ext}\approx B_*$ in the millisecond radio pulsars [unless accretion leads to significant dissipation of the field \citep{konar1, konar}]. This would lead to larger spin-down rates than those observed. On the other hand, if the field remains buried and the magnetic mountain is
stable on long timescales [as simulations by \citet{VIG09} indicate], then the results of \citet{Max1} imply a
quadrupole $Q_{22}>10^{36} (B_*/10^{11}\mbox{G})^{4/3} $ g
cm$^2$ for $M_a=M_c$. From equation (\ref{spindown}), this gives a spin down rate $\dot{\nu}>10^{-14}$ Hz/s for a 500 Hz pulsar, close to
the maximum spin down rates measured for millisecond pulsars. High
initial fields of $B_*\gtrsim 10^{11}$ G would thus challenge current observations.

\subsection{Gravitational radiation}

In the left panel of figure \ref{magfigs1} we plot GW strain vs frequency for magnetic mountains that do not decay between outbursts. We consider first a scenario in which $B_*\approx 10^{10}$ G and the critical mass $M_c$ has been accreted over a system's lifetime (i.e. $M_a=M_c$). The GW emission is predictably weak. Given the
uncertainties associated with modelling field burial, in figure \ref{magfigs1} we also consider the
case in which the birth field is $B_*\approx 10^{12}$ G. In this case some of the
systems could be emitting detectable gravitational radiation, and a detection would provide
evidence for a high degree of field burial.  The latter scenario can
be excluded in the three systems [SAX J1808.4--3658, XTE J1751--305, IGR
J00291+5934 \citep{review}] for which we have a measured spin down between
outbursts. In all cases the spin down rate is
$\dot{\nu}\approx -10^{-15}$ Hz/s and it implies an upper limit of $Q_{22}\approx 10^{36}$ g cm$^2$, if we assume that GW emission is the dominant spin-down mechanism. GW emission at
this level, due to a magnetic mountain, implies $B_*\approx
5\times 10^{10}$ G and would be unlikely to be detected, as can be seen from figure \ref{magfigs1}. For our models  with $M_a=M_c$, $B_*\approx
5\times 10^{10}$ G leads to  $B_{ext}\approx 2.5 \times 10^{10}$ G \citep{Max1}. Such a strong dipole field would, however, lead to a greater
than observed spin down due to magnetic dipole radiation. In fact, if the spin down is attributed to dipole radiation, the implied
magnetic field is $B_{ext}\approx 10^8$ G for all these systems \citep{review}.
An upper limit of $|\dot{\nu}|<2\times 10^{-15}$ Hz/s also exists on
the spin-down rate of Swift J1756.9-2508. In this case the limit on
the dipole field from electromagnetic spin down is of $B_{ext}\lesssim
5\times 10^8$ G but the field needed to explain the spin down in terms
of GWs from magnetic mountains is $B_*\approx 10^{12}$ G corresponding to $B_{ext}\approx 5\times 10^{11}$ G for $M_a=M_c$ in our models.
 It is important to note though that while simulations indicate that the quadrupole saturates for $M_a\gtrsim M_c$ \citep{Wette}, no such effect is observed for the decay of the external field, and the limits on evolving the field further are mainly numerical. One cannot thus exclude high degrees of field burial. In fact the harmonic content of thermo-nuclear bursts suggests that in some systems burning occurs in patches and is confined by locally strong and compressed magnetic fields \citep{BS06,M10,Cavecchi, Chaka}.

We also analyse the scenario in which the magnetic mountain decays on
short timescales between accretion outbursts. Time-dependent MHD simulations show
that magnetic line tying at the stellar surface stabilises the
mountain against interchange instabilities. Current-driven Parker-type
instabilities do occur, but they do not disrupt the mountain,
saturating in a state where the quadrupole is reduced by $\lesssim 60\%$
\citep{VIG09}. Simulations confirm stability up to the
tearing-mode timescales but they do not resolve slower instabilities
and modes below the grid-scale. Different choices of boundary
conditions can also destabilise the system \citep{Dipankar,Muk3}.  In this
scenario we take $M_a=\Delta t \langle \dot{M} \rangle$ for each system, and calculate the quadrupole from equation (\ref{quad}). The results for the predicted gravitational wave strain are shown in the right panel of figure \ref{magfigs1}. This scenario leads
to small mountains and weak GW emission, that would be undetectable for most systems, even for ET. The only systems that would lead to detectable GWs are the persistent ones, if $B_*\approx10^{12}$ G.

\begin{figure}
\centerline{\includegraphics[width=8.5cm]{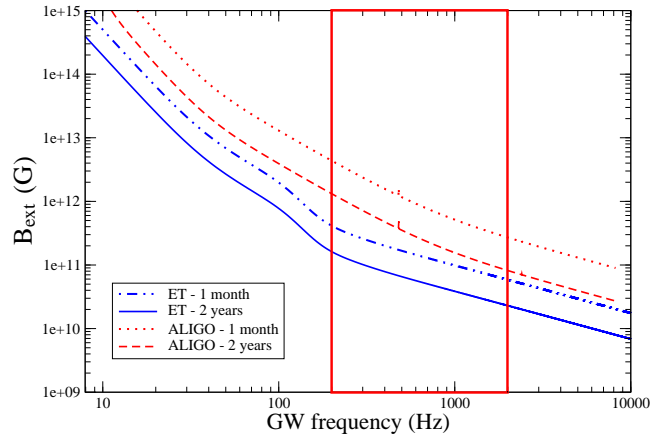}}
\caption{The sensitivity of Advanced LIGO and ET to a magnetic mountain. The GW strain is expressed in terms of the magnetic field $B_{ext}$ of the star, for a fiducial system at 5 kpc and model E of \citet{Max1}. We take $M_a=M_c$ and, as described in the text, one has $B_{ext}=B_*/2$ for these models. We plot both the case of a 1 month integration (dot-dashed curve for ET and dotted curve for Advanced LIGO) and a 2 year integration (solid curve for ET and dashed curve for Advanced LIGO).  We can see Advanced LIGO will probe high field scenarios, with $10^{11}\mbox{ G}\lesssim B_{ext}\lesssim 10^{12}$ G, while ET will be able to probe fields of $B_{ext}<10^{11}$ G. }\label{magfigs}
\end{figure}

In figure \ref{magfigs} we show the GW strain expressed in terms of an equivalent $B_{ext}$ obtained from equation \ref{quad}, using model E of \citet{Max1} and $M_a=M_c$, for which $B_{ext}=B_*/2$. We can see that Advanced LIGO is expected to probe high field scenarios, with $10^{11}$ G $\lesssim B_{*} \lesssim 10^{12}$ G, while ET will probe a physically more realistic section of parameter space, i.e. $B_{*}<10^{11}$ G.



\subsection{Distinguishing magnetic from thermal mountains} 
\label{dist}
An interesting question is if, given a GW detection, it would be possible to understand whether we are observing a thermal or magnetic mountain. We have already discussed the electromagnetic counterpart of a thermal mountain in section \ref{therm}, and showed that a quadrupolar deformation could lead to flux modulations and pulsations in quiescence at twice the spin frequency. The results of the previous section suggest that if a magnetic mountain were to be detected in a hypothetical system, such a NS would have a strong 'birth' (i.e. at the onset of the LMCB phase) magnetic field $B_*\approx 10^{12}$ G, although the external dipolar field may be lower, due to accretion induced magnetic burial. In such a circumstance cyclotron resonance scattering features should appear in the X-ray emission and  \citet{Priymak14} have studied the problem in detail for the case of an accretion buried field. 

\begin{figure}
\centerline{\includegraphics[angle=270,width=9cm]{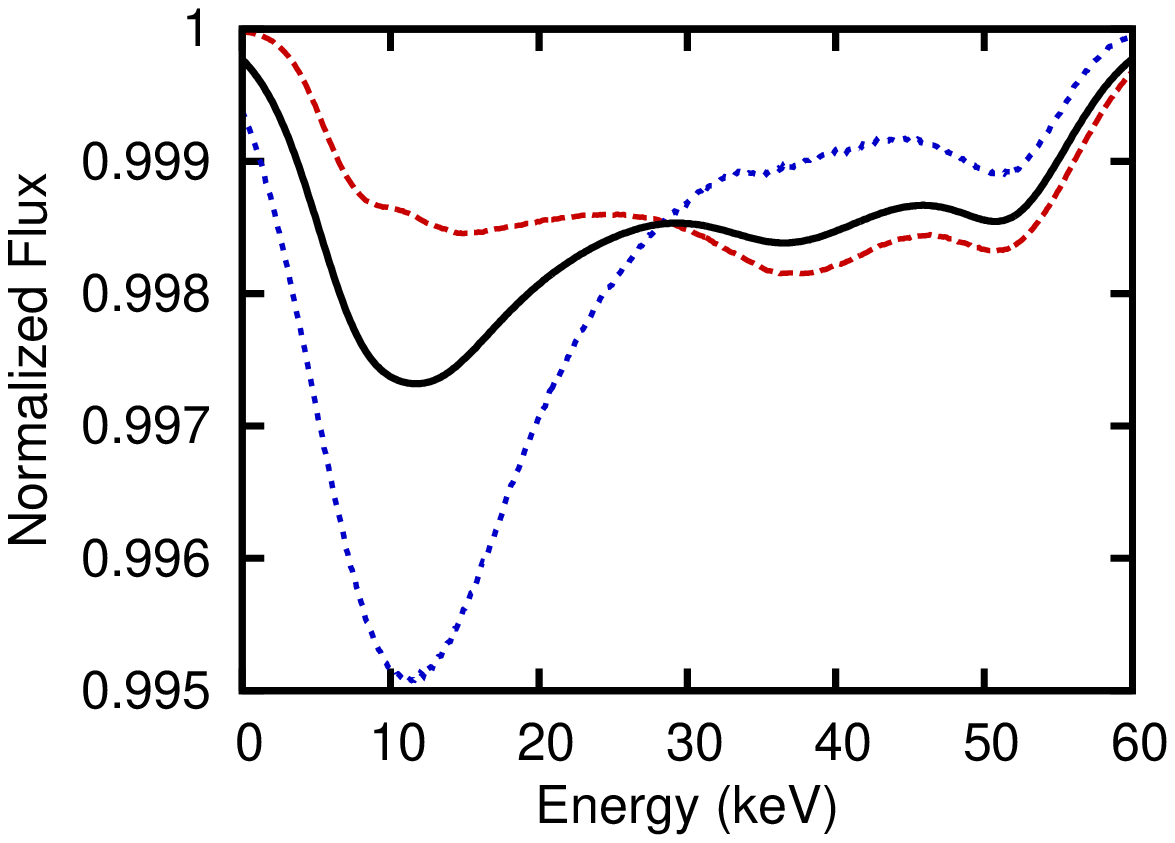}}
\caption{Example of a cyclotron spectrum, obtained with the code of \citet{Priymak14} for a $M=1.4 M_\odot$ NS described by EOS E, with the following parameters: $\iota=\pi /4$ (observer inclination relative to the rotation axis), $\alpha=\pi /4$ (inclination of the magnetic axis relative to the rotation axis), $B_*=10^{12.8}$ G, $M_a=M_c=3.01426\times 10^{-7} M_\odot$ (see \citet{Priymak14} for a full description of the parameters). The solid line represents the phase averaged spectrum, while the dashed and dotted lines represent the phase resolved spectra for two extreme rotational phases, $\omega =\pi /2$ (dashed line) and $\omega =3\pi /2$ (dotted line). While the energy of the lines remains fairly constant the depth varies significantly with phase. The flux is normalised to give unit peak flux.}\label{max2}
\end{figure}

\begin{figure*}
\centerline{\includegraphics[angle=270,width=9cm]{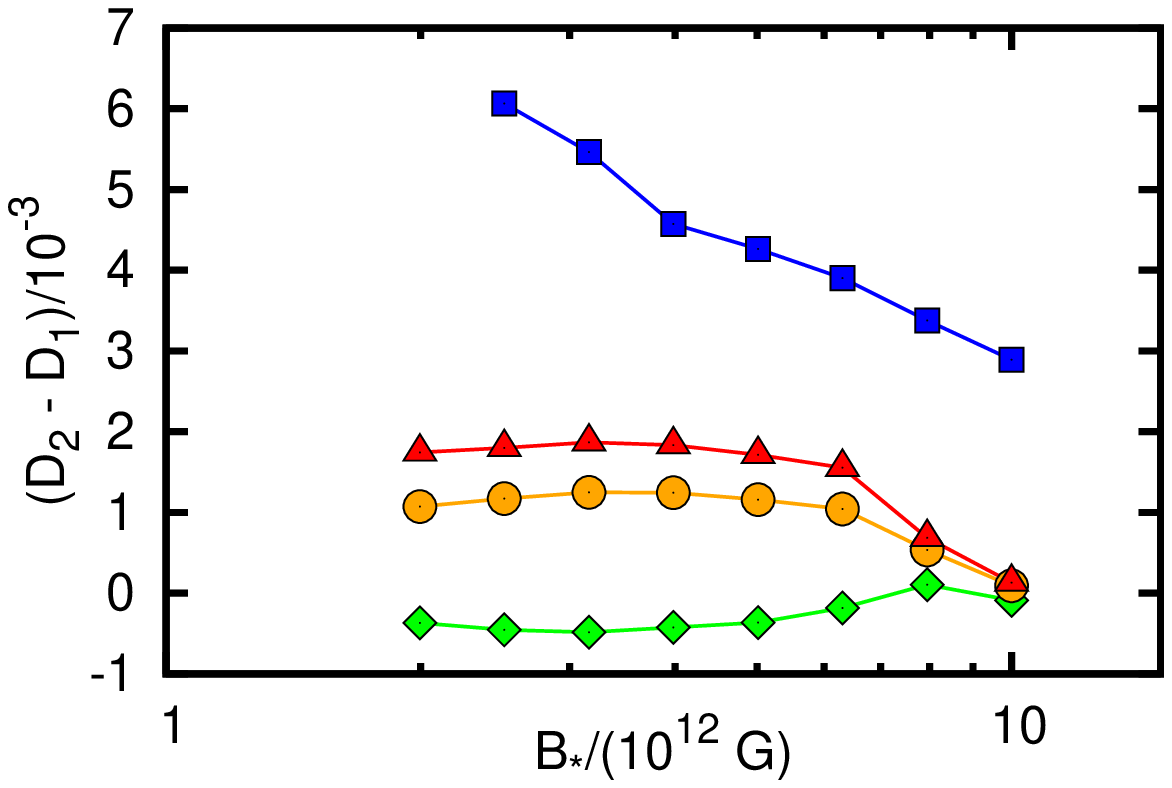}\;\includegraphics[angle=270,width=9cm]{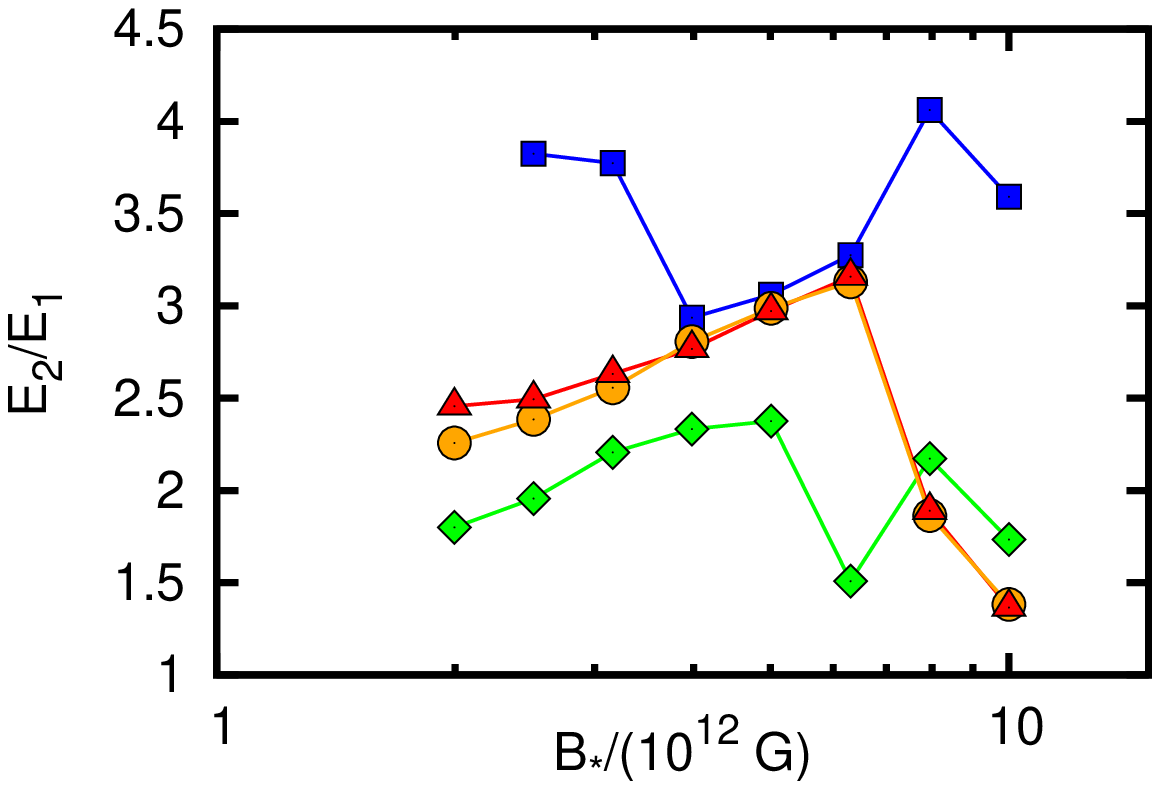}}
\caption{Difference in depth $D$ in normalised flux units (left panel) and ratio between the energies (right panel) of the second and first cyclotron line for a $1.4 M_\odot$ NS obtained with model E of \citet{Priymak14}, for varying field strengths $B_*$. The different colours represent different inclinations of the observer ($\iota$) and of the magnetic filed axis ($\alpha$) with respect to the rotational axis: $\iota=0, \alpha=0$ (blue squares), $\iota=0, \alpha=0.5\pi$ (green diamonds), $\iota=0.5\pi, \alpha=0.5\pi$ (red triangles), and $\iota=0.25\pi, \alpha=0.25\pi$ (orange circles). For $B_*<10^{12} $ the features cannot be distinguished. }\label{max1a}
\end{figure*}
We repeat the analysis here for a 1.4 $M_\odot$ NS with an accreted outer envelope described
by equation of state E of \citet{Priymak14}. We
vary  $B_*$ between $10^{11}$ G and
$10^{12}$ G and study the emission features for $M_a=M_c$.  In figure \ref{max2} we show an
example of the kind of spectra that such a setup produces. The
solid line represents the phase averaged spectrum, while the dotted
lines represent phase resolved spectra for two extreme rotational
phases, $\omega=\pi/2$ and $\omega=3\pi/2$. We can see that in all
cases the energy of the first line is fairly stable, but the depth can
vary strongly with phase, as can the shape of the higher energy
features. A strong phase dependence of the fundamental line for different sizes of polar mountains has also been found by \citet{Muk2}.

Let us focus on the phase averaged spectrum. Our simulations show that for $B_*\lesssim 10^{12}$ G no
cyclotron resonance scattering features are present. The results for higher field strengths are shown in figure
(\ref{max1a}), where we plot the difference in depth between the first
and second line and the ratio between the energies at which the lines
appear, versus the pre-accretion magnetic field $B_*$. The effects are small, but may be measurable by
future X-ray observatories such as NICER and LOFT, which will both be
capable of resolving modulations of less than $1\%$ at energies of
$\approx 1$ keV \citep{future1, future2}. Furthermore the cyclotron features appear to be more pronounced
in the region of interest, i.e. the field strengths that would lead to
GW emission at the Advanced LIGO and ET threshold. This method thus
has the potential to be a good diagnostic for distinguishing different
kinds of continuous GW emission.  Additionally instruments such as
NICER and LOFT will also be able to carry out phase resolved
spectroscopy, allowing for a much more detailed
characterisation of the cyclotron resonance scattering features in
these systems, which can vary significantly with phase, as
illustrated in figure \ref{max2}.


We stress here that no cyclotron lines have been detected in LMXBs
containing neutron stars rotating with millisecond periods, and the fields of these systems
are generally thought to be reasonably weak ($B_{ext}\approx 10^8$
G). Nevertheless if an as yet unobserved system (e.g. a system that is
currently in quiescence) were to become visible and emit
detectable GWs, the presence of a cyclotron line would point to a
magnetic mountain. Its absence, on the other hand, combined with the
estimates in section \ref{magfigs}, would suggest that the quadrupole 
is more likely to be due to thermally generated crustal mountain [although
mountains in the core of the star are also a possibility \citep{CFLm}]. Especially for weaker fields, however, several combinations of
orientation and inclination could lead to cyclotron resonance scattering features not being
detectable (see \citet{Priymak14} for an in depth discussion), so their absence is inconclusive.

\section{Conclusions}

In this paper we asses the likely gravitational wave signal
strength and detection prospects for deformations, or `mountains' on
neutron stars in LMXBs. Unlike most previous work on this topic we do not
assume that the gravitational wave spin-down torque has to balance the
accretion induced spin-up torque, as several studies have indicated that
this is unlikely to be the case for many systems \citep{A1,BrynAle,ABCPHD}. Rather, we
calculate the gravitational wave signal strength due to the two
main mechanisms that have been suggested for building a
mountain: asymmetric thermal deposition in the
crust (thermal mountains) and magnetically confined mountains
(magnetic mountains).  We calculate the gravitational wave
strain for both mechanisms in known LMXBs for which we can
measure the spin frequency, average accretion rate during outbursts and
outburst duration.

One of the main uncertainties is the timescale on
which the mountain is stable once accretion ceases and the system enters quiescence. For thermal
mountains it is likely that the quadrupole will dissipate on a thermal timescale $\tau_{th}\lesssim 6 $ years, leading to
large mountains only in persistently accreting systems. 
In this scenario the GW signal strength for most
transient systems falls below the level that would be detectable by
Advanced LIGO or ET. In the case of persistent systems, however, the mountain could be even larger than what is
required for torque balance, if the crust is as strong as
predicted by simulations \citep{chuck}. This would not only lead to detectable
gravitational waves, but also predicts a spin-down rate of the neutron star that could be measurable if accretion powered pulsations were to be discovered from these systems.
Continued deep searches for pulsations from luminous LMXBs, such as Sco X-1, are thus complimentary to ongoing GW searches from these systems \citep{sammut14} and could provide crucial constraints.

For the magnetic case simulations indicate that the mountain could be
stable on long timescales (essentially the Ohmic dissipation timescale
$\tau_o\approx 10^8$ years), building up over multiple accretion outbursts. The size of the mountain is strongly dependent on
the strength of the magnetic field when accretion begins, $B_*$. This
is  not well constrained, but the systems we consider are
 old systems, in which the magnetic field is thought to have decayed and to be weak. For both LMXBs and millisecond radio pulsars (that are expected to form mostly from LMXBs) the inferred {\it exterior} field strengths are 
$B_{ext}\approx 10^8$ G.  The exterior dipolar field will, however, be
quenched as the magnetic field is buried by accretion. Our
simulations suggest that the exterior field will be reduced by
approximately two orders of magnitude \citep{Andrew2, Max1}, but the process does not appear to saturate, and the limits on pushing the results further are mainly numerical. Further field burial is thus possible.  We consider two
scenarios: one in which $B_*=10^{10}$ G and the other in which $B_*=10^{12}$ G.
For $B_*\approx 10^{10}$G  the detection
prospects for magnetic mountains are pessimistic. For a detection with
Advanced LIGO or ET it is necessary to have an initial magnetic field
 $B_*\approx 10^{12}$ G. Although this appears unlikely
for currently observed LMXB systems, for which the evidence suggests
weakly magnetised neutron stars \citep{dangelo}, the process
of magnetic burial is still not well understood, and such high values
of the background field cannot be excluded.

Finally, it is interesting to
note that if a mountain is detected by LIGO or ET, it could be possible
to distinguish between a thermal or a magnetic mountain.  For the
relatively high values of the magnetic field $B_*\approx 10^{12}$ G that make the
magnetic mountain detectable one would, in fact, expect phase-dependent and non-trivial cyclotron
resonance scattering features to be present in the X-ray spectrum. We
calculate examples of such features and show that they
could be detected by future X-ray observatories, such as LOFT or
NICER. A detection of a gravitational wave signal combined with a
detection of cyclotron features would provide a strong direct indication of a
magnetic mountain and of a large buried magnetic field.

\section*{Acknowledgments}

BH acknowledges the support of the Australian Research Council (ARC) via a Discovery Early Career Researcher Award (DECRA) fellowship. This work is also supported by an ARC Discovery Project grant.

\end{document}